\documentclass[aps,showpacs,prb,floatfix,twocolumn]{revtex4}
\usepackage{amsmath,amssymb,graphicx,bm,epsfig,color}

\newcommand{\vk}{{\mathbf{k}}}

\begin{document}

\title{Screening of magnetic moments in PuAm alloy : LDA+DMFT study}

\author{J. H. Shim$^1$, K. Haule$^1$, S. Savrasov$^2$,  and G. Kotliar$^1$}

\affiliation{$^1$ Department of Physics, Rutgers University, Piscataway,  NJ 08854, USA}
\affiliation{$^2$ Department of Physics, University of California, Davis, CA 95616}

\begin{abstract}
  The puzzling absence of Pu magnetic moments in a PuAm environment is
  explored using the self-consistent Dynamical Mean Field Theory (DMFT)
  calculations in combination with the Local Density Approximation. We
  argue that $\delta$-Pu -Am alloys provide an ideal test bed for
  investigating the screening of moments from the single impurity
  limit to the dense limit. Several important effects can be studied:
  volume expansion, shift of the bare Pu on-site $f$ energy level, and
  the reduction of the hybridization cloud resulting from the
  collective character of the Kondo effect in the Anderson lattice.
  These effects compensate each other and result in a coherence scale,
  which is independent of alloy composition, and is around 800$\,$K.
  We emphasize the role of the DMFT self-consistency condition, and
  multiplet splittings in Pu and Am atoms, in order to capture the
  correct value of the coherence scale in the alloy.
%This study highlight the
%  importance of mixed valence in this region of the actinide series.
\end{abstract}
\date{\today}
\maketitle

The electronic structure of plutonium (Pu) is one of the outstanding
issues in condensed matter theory. The volume of $\delta$-Pu is
between that of the early actinides, where the $f$ electrons are
itinerant, and the late actinides, where the $f$ electrons are
localized. Hence the view that the $f$ electrons are very close to a
localization-delocalization transition was put forward early on by
Johansson~\cite{Johansson}.

Theoretical density functional theory calculations in many
implementations, all predict $\delta$-Pu to be magnetic\cite{lda_pu}.
%with a static moment of the order of Bohr magnetons. 
%
%SORRY HOW MANY???
%
Nevertheless, various experiments that searched for the magnetic
moment in $\delta$-Pu, such as specific heat in field, neutron
scattering and $\mu$SR measurements\cite{Lashley03,Heffner06}, found
no evidence for the ordered, or fluctuating moment in the energy and
temperature windows probed by these experiments.
High energy spectroscopies, such as electron-energy-loss spectroscopy
and X-ray absorption spectroscopy experiments of the $5d$-$5f$
core-valence transition, clearly indicate that $f$ electrons in Pu are
in a 5$f^5$ configuration\cite{XAS}, which carries a magnetic moment.
Reconciling these low energy and high energy experiments, has lead to
numerous conflicting theories.  Within the most recent Dynamical Mean
Field Theory (DMFT) studies of $\delta$-Pu, allowing for magnetic and
non magnetic solutions, Pu does not order magnetically, and instead
forms a non-magnetic Fermi liquid with a coherence temperature $T^*$
of the order of 800$\,$K\cite{Shim07}.  
%
% The physical picture that
% emerged from that study is a complete compensation of the magnetic
% moment mostly by $spd$ electrons and to some degree by the $f$
% electrons on neighboring atoms. 
%
The Pu atoms are in a mixed valent state, with a dominant $5f^5$
configuration, but with sizeable admixture of $5f^6$.  On the basis of
this picture it was expected, and shown explicitly via a landmark
continuous-time quantum Monte Carlo calculation \cite{ctqmc}, that
expanding the lattice would reduce the coherence temperature $T^*$,
making the moment of the $f^5$ configuration observable.  Other
theories, whereby the $f$ moments are quenched directly by spin
pairing, which is mediated by the spin orbit coupling, also account
for the absence of moments in Pu, and suggest that this material is at
a quantum critical point\cite{Chapline07}.

\begin{figure}[bt]
\includegraphics[width=0.6\linewidth]{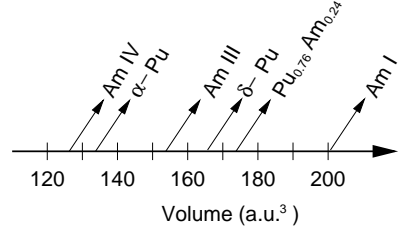}
\caption{ 
The volume of Pu and Am in each phases.  According to
Ref.~\onlinecite{Baclet07}, the volume of Pu$_{0.76}$Am$_{0.24}$ is
locates slightly above ($\sim 2\%$) the value estimated from the
linear interpolation between $\delta$-Pu and Am-I.
%
%What is PuAm? Is it Pu$_{0.76}$Am$_{0.24}$. CHANGE $\beta$-Pu to
%$\delta-$Pu.  
}
\label{fig1}
\end{figure}

PuAm alloy is a unique system to study the effect of lattice expansion
of $\delta$-Pu since the volume of Americium at ambient pressure
(Am-I) is 20\% bigger than volume of $\delta-$Pu (see
Fig.~\ref{fig1}). Moreover, there is no structural phase transition
between pure $\delta-$Pu and Pu$_{0.25}$Am$_{0.75}$, and the structure
remains $fcc$\cite{Ellinger66}.
Surprisingly, an intensive program studying the Pu-Am mixtures, which
has the effect of expanding the distance between the Pu atoms, has not
shown an enhanced magnetic susceptibility, nor a narrowing of the peak
at the Fermi level, as measured by photoemission
experiment~\cite{Javorsky06,Baclet07}.  This seems naively at odds
with both the quantum critical picture, and the DMFT picture, and
supports the suggestion, that the configuration of Pu in the solid is
close to an inert $5f^6$ configuration \cite{Shick05,Pourovskii06},
being essentially equivalent to Am configuration.  Unfortunately, this
picture can not explain the results of high energy
spectroscopies\cite{XAS}, nor the difference in specific heat between
$\alpha$-Pu and $\delta$-Pu\cite{Shim07,Pourovskii07}, nor the
spectral properties\cite{Zhu07}. Hence the puzzle of the absence of
magnetic moments in Pu-Am alloys remains.

In this paper, we provide a theoretical explanation of this puzzle
using LDA+DMFT method \cite{rmp_georges,rmp_kotliar}.  We find that indeed the
moment remains well screened in agreement with experiment, and we
identify the mechanism that compensates the effects of the lattice
expansion.  While the Pu-Am system cannot realize the unscreening of
the local moment, and hence the approach to quantum criticality, it
provides an ideal system in which to follow the transition from
collective screening of moments, to single impurity spin screening of
moments.
%
% We use this system to elucidate the role of the DMFT self-consistency
% condition on the effective hybridization function, and how it differs
% from the bare hybridization function, and predict a marked increase in
% resistivity upon alloying and a disappearance of the optical
% hybridization gap, which should be present in pure Pu metal.
 
%{\bf  Jihoon,  we never published optics of Pu. Perhaps we should
%have a separate article illustrating this } 

%To address the issues of Pu-Am alloy, we use dynamical mean field
%theory (DMFT)\cite{rmp_georges} in combination with the local density
%approximation (LDA+DMFT)\cite{rmp_kotliar}.  

Within the LDA+DMFT method\cite{rmp_kotliar}, the itinerant $spd$
electrons are treated by the LDA method\cite{Savrasov96}, while for
the electrons in the correlated $f$ orbitals, all the local Feynman
diagrams are summed up by solving an auxiliary quantum impurity
problem, subject to a DMFT self-consistency condition
\begin{equation}
\frac{1}{\omega+\mu-E_{imp}-\Sigma-\Delta}=\sum_\vk \left[{G_{\vk}}(\omega)\right]_{ff}.
\label{Eq1}
\end{equation}
Here $\Delta(\omega)$ is the hybridization matrix of the auxiliary
impurity problem, $E_{imp}$ is the matrix of impurity levels, and
$G_\vk$ is the one electron Green's function
$G_\vk=1/(O_\vk(\omega+\mu)-H_\vk-\Sigma)$. $H_\vk$ and $O_\vk$ are
the LDA Hamiltonian and the overlap matrix in the projective
orthogonalized LMTO basis set \cite{Toropova07}. The quantum impurity
solver is used to determine the local self-energy correction to the
$f$-orbital Hamiltonian, which is a functional of the impurity levels,
and the hybridization strength $\Sigma[E_{imp},\Delta]$.
To solve the impurity problem, we used the vertex corrected
one-crossing approximation method~\cite{SUNCA}, and the results are
further cross-checked by the continuous time quantum Monte Carlo
method\cite{Haule07}. The Slater integrals $F^k$ ($k$ = 2, 4, 6) are
computed by Cowan's atomic Hartree-Fock program including relativistic
corrections\cite{Cowan} and rescaled to 80\% of their atomic value,
accounting for the screening in the solid. We use Coulomb interaction
$U=4.5\,$eV, consistent with previous studies of Pu and
Am\cite{Savrasov01,Savrasov06}.

The Pu$_{1-x}$Am$_x$ alloys are approximated by the charge ordered
compounds with $x$ = 0, 1/4, 1/2, and 3/4. Based on the original $fcc$
phase, we use a four times larger unit cell with the simple cubic
structure. To determine the lattice constant of the alloy, we used the
linear interpolation between the two end compounds, $\delta$-Pu and
$\beta$-Am\cite{Baclet07}. We use 8$\times$8$\times$8 momentum points
in the first Brillouin zone in combination with the tetrahedron
method.

\begin{figure}[tb]
\includegraphics[width=1.0\linewidth]{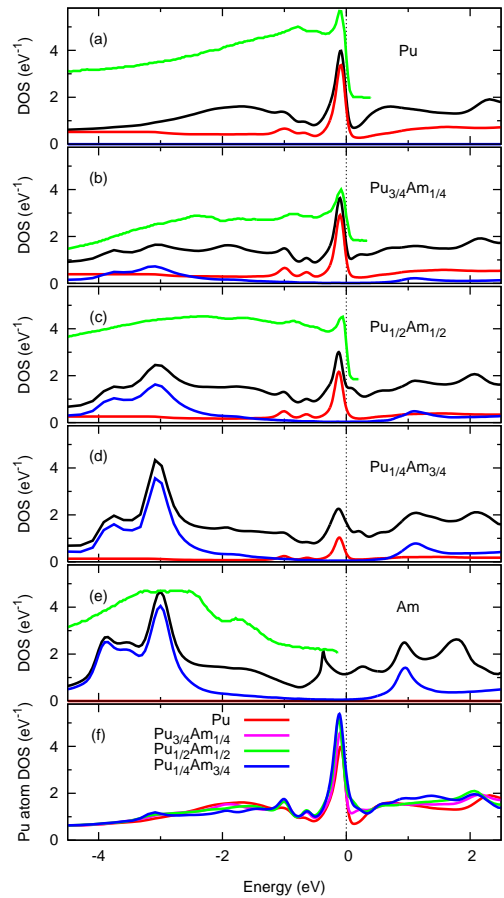}
\caption{
(a-e) The spectral function of Pu$_{1-x}$Am$_{x}$ in the $fcc$ unit cell.
Total (black), Pu $5f$ (red), Am $5f$ (blue) spectrum are obtained by the
LDA+DMFT method.
The green lines show experimental photoemission spectrum at 
$x$ = 0.0 (a), 0.20 (b), 0.36 (c), and 1.0 (e) 
(extracted from Ref.~\onlinecite{Baclet07}). 
(f) The spectral function of Pu atom for various Am dopings.
}
\label{fig2}
\end{figure}

Figure \ref{fig2} (a-e) shows the calculated spectral function of
Pu$_{1-x}$Am$_{x}$. The theoretical results are compared to the
experimental spectra of $\delta$-Pu alloyed with Am\cite{Baclet07}. 
Naively, one would expect that alloys of Pu and Am are magnetic,
because the volume of the alloy is bigger than the volume of the
elemental plutonium. Increasing the volume decreases the hybridization
between the $5f$-orbital and the conduction spd-orbitals, and
consequentially the Kondo coupling $J$ should be reduced.  Since the
strength of the Kondo screening depends exponentially on the Kondo
coupling $J$ while RKKY scale is only a quadratic function of $J$, one
is expecting the screening to become inefficient even for a small
change of $J$.
However, experiments ruled out the possibility of magnetic moment
formation.
%
%In alloyed cases, people expected unscreening of magnetic moments,
%because the $J_{Kondo}$ decrease much faster than $J_{RKKY}$ as
%increasing the Pu-Pu distances.  However, theses features are not
%observed in experiments.  
%
The specific heat of Pu$_{1-x}$Am$_{x}$ does not show any significant
changes\cite{Javorsky06} with alloying and is simply proportional to
the Pu content. This is because the inert $f^6$ configuration of Am
has very small specific heat, and only Pu atoms contribute to $c_v$.
The photoemission spectral of Pu$_{1-x}$Am$_{x}$ has been reproduced
by the simple summation of the spectrum of pure $\delta$-Pu and
Am\cite{Baclet07}.

Our LDA+DMFT results in Fig.~\ref{fig2} show a very favorable
agreement with experiment. 
The spectral function of Pu atom, displayed in Fig.~\ref{fig2} (f), is
almost unchanged in PuAm alloys.
%not much effected by alloying with Am, when the volume change is taken
%into account. 
The coherence scale, that determine the temperature below which the Pu
moment is screened, remains roughly at 800$\,$K in the alloys.
The Am spectra is even more doping independent and shows no appreciate
change when alloyed with Pu. This is not surprising since a very large
pressure is needed to delocalize Am $5f$ electrons
\cite{Heathman00,Savrasov06}, and only the Am IV phase is known to be
itinerant (see Fig.1 for location of Am IV in volume diagram).
corresponding to volume of Am IV in Fig.
\ref{fig1}.
Taking into account that the spectra of both Pu and Am are almost
unchanged in alloys, the total DOS can be well approximated by a
linear combination of the DOS of the two end compounds, as conjectured
in Ref.~\onlinecite{Baclet07}.

% The mysterious absence of the magnetic moments in Pu-Am alloys can be
% traced back to the self-energy correction, $\Sigma_{Pu f}$, on Pu
% site, which is almost independent of alloy composition. Since the
% volume is considerably increased by Am alloying, the hybridization
% strength $\Delta$ is suspected to decrease. Since the coherence scale
% is exponentially sensitive to the hybridization strength (and impurity
% levels), even a tiny change in hybridization could result in
% substantial change of coherence scale and hence appearance of
% unscreened magnetic moments.

In order to gain further insight into the puzzle of the doping
independent coherence scale in Pu$_{1-x}$Am$_{x}$ compound, we will
examine the important factors that govern the screening of the
magnetic moments in our LDA+DMFT calculation. 
The temperature dependence of the spectra and hence the coherence
scale is entirely determined by the DMFT self-energy correction
$\Sigma$, which is a functional of the hybridization function
$\Delta$, impurity levels $E_{imp}$ and the screened Coulomb
interaction $U$, i.e., $\Sigma[\Delta, E_{imp},U]$. $U$ is here kept
constant. The hybridization function $\Delta$ and impurity levels
$E_{imp}$, determined by Eq.~\ref{Eq1}, are both quite sensitive to
the volume change as well as chemical substitutions, as we will show
below.

The coherence scale can be expressed analytically when the following
two assumptions are made: a) The atom has SU(N) symmetry and hence the
spin-orbit, crystal field splittings and Hund's coupling are absent, b)
the hybridization function $\Delta$ is frequency and temperature
independent. In this case, the Kondo coherence scale 
% 
% depends exponentially in the
% hybridization and this exponential dependence 
% 
can be estimated by $T_K \sim e^{-1/J}$, where $J$ is
\begin{equation}
J \sim \frac{N}{\pi}\left(\frac{|\Delta''|}{E_{n+1}-E_n}+\frac{|\Delta''|}{E_n-E_{n-1}}\right).
\label{Eq2}
\end{equation}
and $E_n$ are the atomic energies corresponding to occupancy $n$, $N$
is degeneracy of the SU(N) model. In the extreme Kondo limit, the
difference in the atomic energies are $E_{n+1}-E_{n}=U/2$ and
$E_{n}-E_{n-1}=U/2$.
The above assumptions are not realistic in Pu because of the strong
multiplet effects and the strong frequency and temperature dependence
of the hybridization function $\Delta(\omega,T)$. Nevertheless, the
formula Eq.~\ref{Eq2} gives the hint that the coherence scale is
sensitive to the ratio $\langle |\Delta''|\rangle/\Delta E_f$, where
$\langle |\Delta''|\rangle$ is average of the hybridization function
in some low energy region, and $\Delta E_f$ is the difference between
the ground state of the atomic $f^6$ configuration and the ground
state of the atomic $f^5$ configuration. Namely, Pu $5f$ electrons are
in mixed-valence state with an average occupancy close to $n_f\approx
5.2$ \cite{Shim07} therefore the dominant fluctuations are
\begin{equation}
e + 5f^5 \leftrightarrow 5f^6, 
\end{equation}
where $e$ is a conduction $spd$ electron.

\begin{figure}[tb]
\includegraphics[angle=270,width=0.8\linewidth]{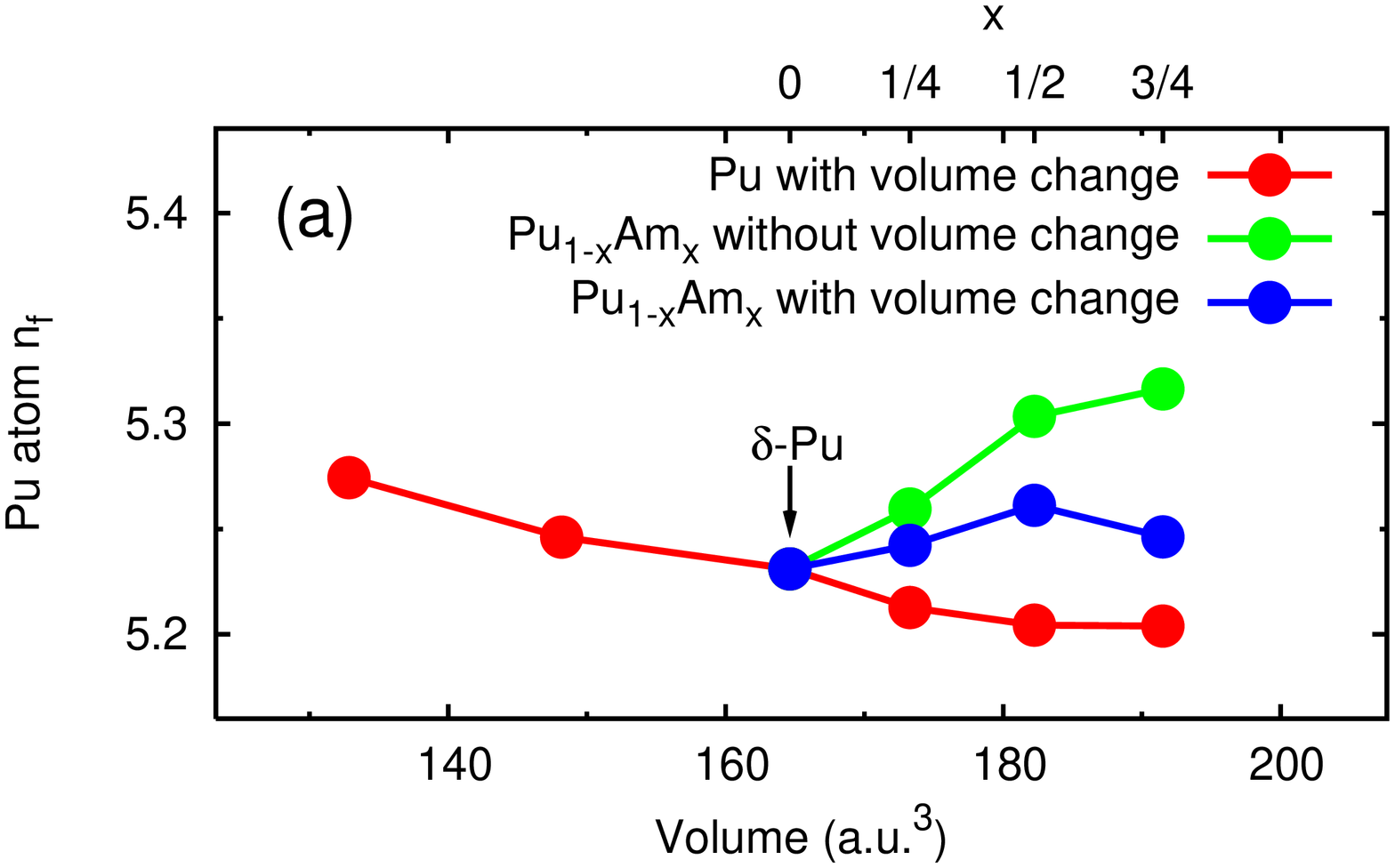}
\includegraphics[angle=270,width=0.8\linewidth]{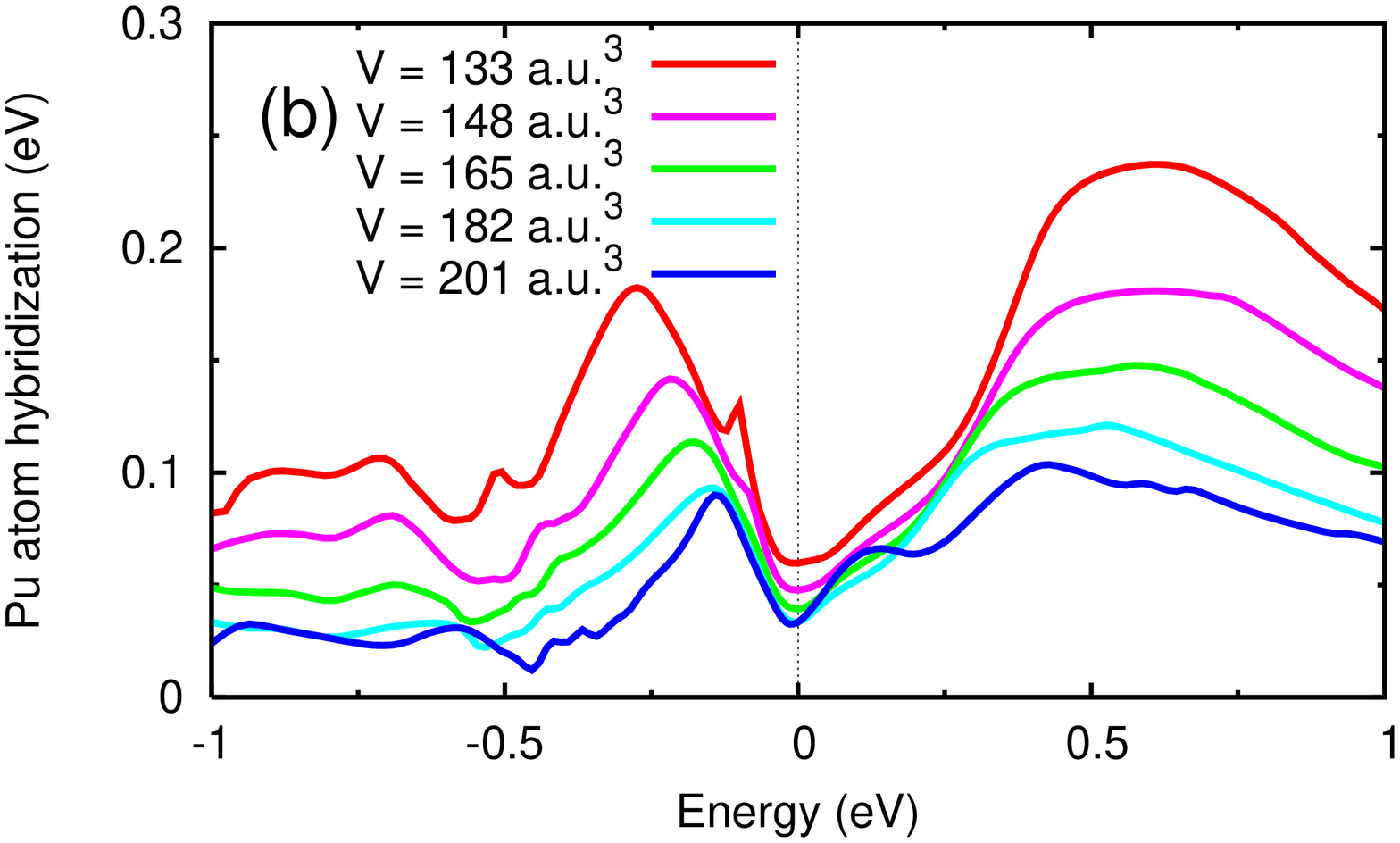}
\includegraphics[angle=270,width=0.8\linewidth]{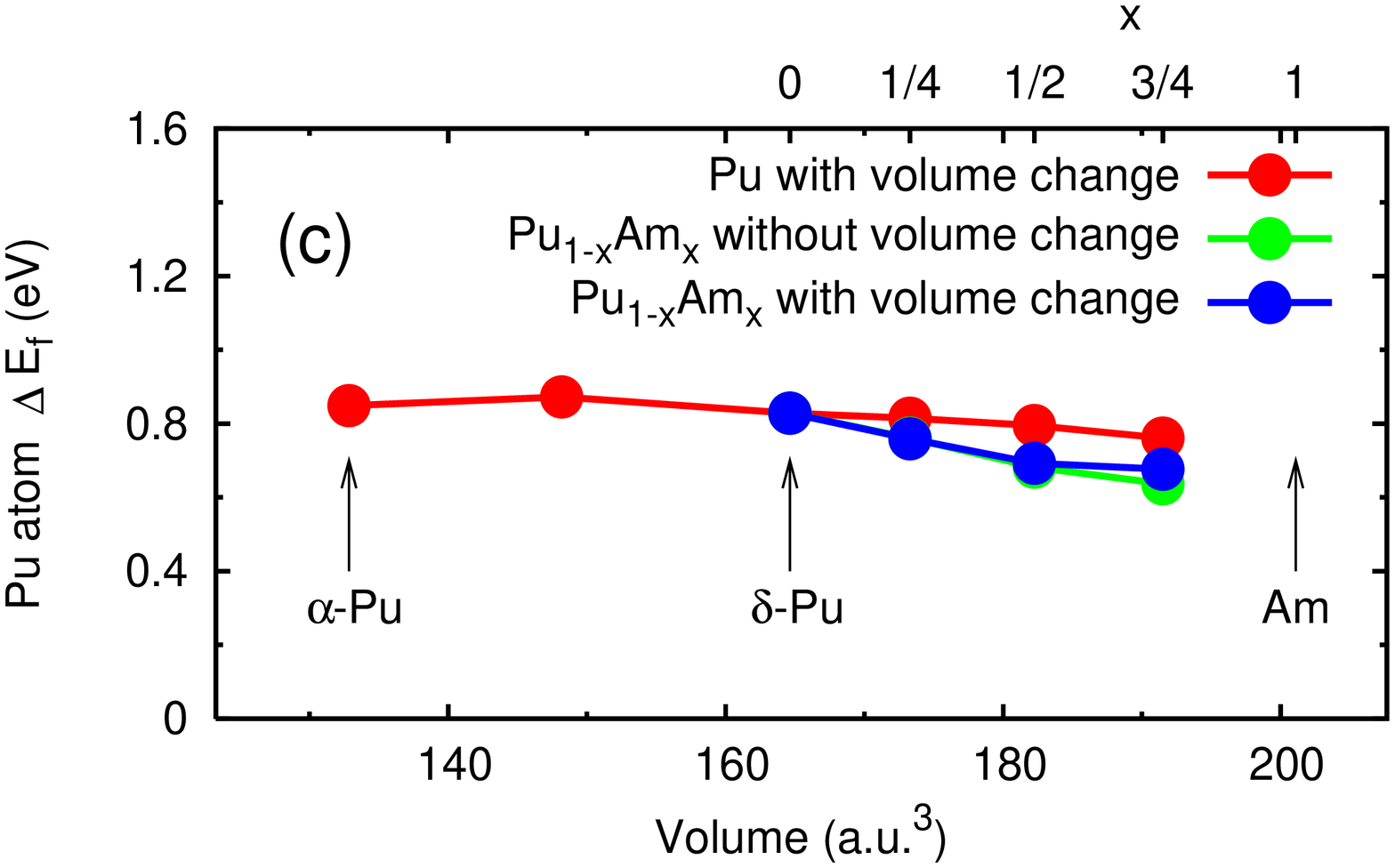}
\caption{ (a) The number of $5f$ electrons, $n_f$, as a function of
  volume (red dots), doping $x$ at constant volume (green dots), and
  combination of both (blue dots).  Note that $a$ = 8.1, 8.7 and 9.3
  a.u. correspond to the volume of $\alpha$-Pu, $\delta$-Pu and
  $\beta$-Am, respectively.  (b) Volume dependence of hybridization
  function for Pu in $fcc$ phase.  (c) The energy difference between
  the two most relevant atomic states, defined in the text.  }
\label{fig3}
\end{figure}
In turn, the $5f$ occupancy strongly cooperates  with the coherence
scale, and the scale is smallest in the Kondo regime where $n_f$ is
integer.
%, and is largest for half-integer occupancies. 
In
Fig.~\ref{fig3} (a) we show the $5f$ occupancy for three different cases
of chemical substitution and volume change. If the volume of the
elemental Pu is changed without alloying, $n_f$ is slightly decreasing
with increasing volume (see red curve in Fig.~\ref{fig3} (a)). This is to
be expected since in the extreme limit of large volume the system will
be in the extreme Kondo limit with $n_f=5$. The magnitude of the $n_f$
variation in the range of volumes, corresponding to $\alpha$-Pu and Am
is however surprisingly small, only of the order of $0.06$.  If the
volume is kept constant, but the the chemical composition is varied,
the occupancy $n_f$ is increasing with Am doping (green dots in
Fig.~\ref{fig3} (a) correspond to $x=0$, $1/4$, $1/2$, and $3/4$ from
left to right). The Pu atom is thus more mixed valent in the limit of
small Pu concentration, i.e., single impurity limit. This might seem
counter-intuitive, but the increase in the coherence scale can be
understood from the following consideration. The quasiparticle peak
detected in $5f$ partial DOS creates a depletion in the conduction spd
bands. This is traditionally called Kondo hole, which is always pinned
at the Fermi level. The hybridization is thus decreased due appearance
of quasiparticle peak, an effect termed by Noziere as ``exhaustion''.
This Kondo hole does not form on Am atom, since there is no coherence
peak on Am atom. Thus, if a single Pu atom in emerged in a lattice of
Am atoms, the moment can be more efficiently screened by
spd-electrons, residing on neighboring Am sites.
Finally the blue dots in Fig.~\ref{fig3} (a) show that the two effects
largely cancel, when both the substitution and volume change are taken
into account, and the $5f$ occupancy of Pu$_{1-x}$Am$_x$ is almost
constant.

The frequency dependent hybridization function
$|\Delta''(\omega)|/\pi$ for elemental Pu at different volumes (no
alloying), is shown in Fig.~\ref{fig3} (b). The Kondo hole, i.e.,
depletion of hybridization at zero frequency, is clearly visible.
Moreover, the hybridization changes substantially when volume is
varied between the volume of $\alpha$-Pu and Am volume. The change is
between 30\% at low energy and up to 50\% at intermediate energies.
This is a substantial change, which indeed unscreens magnetic
moments and changes Pauli susceptibility into Curie-Weiss
susceptibility, as shown in Ref.~\onlinecite{ctqmc}.

The Eq.~\ref{Eq2} reveals that the coherence scale depends sensitively
on the ratio $|\Delta''|/\Delta E_f$ rather than on the hybridization
itself. In Fig.~\ref{fig3} (c) we plot the smallest energy difference
$\Delta E_f=E_6-E_5$, which is the dominant contribution to the
exchange coupling in Eq.~\ref{Eq2}. Here $E_6$ and $E_5$ are the
ground state energies of the atomic $f^6$ and atomic $f^5$ configuration,
respectively.  In Fig.~\ref{fig3} (c) we separate the effect of volume
increase from the chemical substitution. It is clear from
Fig.~\ref{fig3} (c) that the energy difference is almost three times
smaller than $U/2\sim 2.25eV$, the value that corresponds to the
extreme Kondo limit. The $5 f^6$ atomic configuration has thus
substantial overlap with the ground state wave function of the solid,
as shown by valence histogram in Ref.~\onlinecite{Shim07}.  It is also
apparent from Fig.~\ref{fig3} (c) that the volume increase has only a
small effect on this energy difference (red curve). When alloying with
Am, the change of $\Delta E_f$ is larger (blue dots) and is of the
order of 25\%. Since $\Delta E_f$ is decreasing, it thus partially
compensates the decrease of hybridization with pressure, shown in
Fig.~\ref{fig3} (b). However, the magnitude of $\Delta E_f$ variation is
too small to compensate the dramatic change of hybridization.

\begin{figure}[htb]
\includegraphics[angle=270,width=0.9\linewidth]{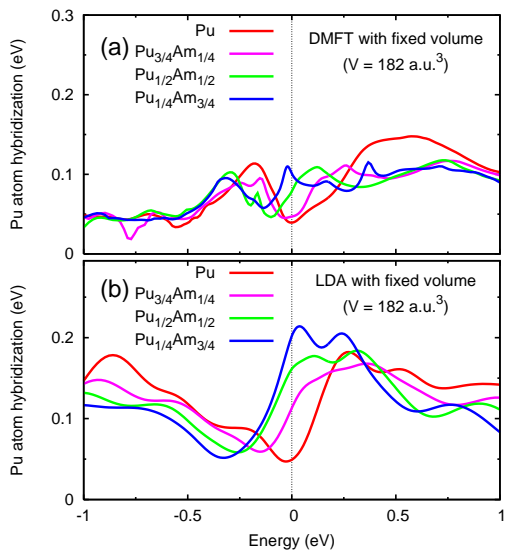}
\caption{
DMFT hybridization function for Pu$_{1-x}$Am$_{x}$ at fixed volume
(a) computed within the self-consistent LDA+DMFT method, (b) within
LDA by setting $\Sigma=0$.
}
\label{fig4}
\end{figure}
%
%Finally, in Fig.~\ref{fig4}a we show the DMFT hybridization function
%for the pure Pu and the three alloy compounds for the case of fixed
%volume. 
%
Finally, Fig.~\ref{fig4}a shows the hybridization function for each
composition of Pu$_{1-x}$Am$_{x}$ alloy for the case of a fixed
volume.  Because the dominant contribution to hybridization function
is not the $f$-$f$ hybridization but the $f$-$spd$
hybridization\cite{Toropova07}, the overall magnitude of the
hybridization is not changed.  However, it is apparent from
Fig.~\ref{fig4} (a) that the chemical substitution of Pu by Am eliminates
the hybridization dip, and therefore the Pu moment is even more
efficiently screened by the spd electrons from neighboring Am atoms
than neighboring Pu atoms.  This elimination of Kondo hole by the
chemical substitution by Am thus compensates the decrease of
hybridization due to volume change. The net change of $\Delta E_f$ and
the change of hybridization $\langle |\Delta''|\rangle$ due to both
volume decrease and alloying compensate and lead to a constant
coherence scale across the phase diagram.

Fig.~\ref{fig4} (b) shows that the increase of hybridization due to
alloying is present even on the LDA level, i.e., computing $\Delta$ by
Eq.~\ref{Eq1} and neglecting the self-energy corrections. However, the
magnitude of the hybridization $\Delta$ and the magnitude of the
change of $\Delta$ is substantially larger in LDA than in
self-consistent LDA+DMFT method. The ``need for renormalization'' of
the LDA hybridization was pointed out very early on
\cite{Gunnarsson,Allen} in the context of the description of cerium
photoemission, when compared to a single impurity calculations. 
%
% This empirical finding can now be put on firm theoretical grounds by
% the DMFT self-consistency condition Eq.~\ref{Eq1}, which determines
% ``the best'' hybridization function for the Pu atom.

To conclude, LDA+DMFT accounts for all the salient features of the
experiments, and give new insights into the mechanism for spin
compensation in the late actinides.  It isolates two essential
elements for the screening of the magnetic moment: hybridization and
position of the $f$ level.  While hybridization is decreased by volume
expansion, as naively expected, the substitution of Am eliminates the
dip in the hybridization function, and at the same time shifts the
position of $f$ level. The three effects compensate each other, and
the coherence scale remains around 800$\,$K. Hence the Pu-Am mixtures
accentuates the mixed valent character of Pu.
%Our work clarifies the
%importance of the DMFT self consistency condition which captures the
%collective character of the screening in lattice models.

Our work has important experimental consequences. While thermodynamic
quantities such as susceptibility and specific heat should not exhibit
significant modifications upon alloying, the same is not true 
for transport quantities.  Optical conductivity is an excellent probe
of the hybridization gap, and we predict that upon alloying, this
hybridization gap will diminish while   the dc resistivity
will rapidly increases.

Acknowledgment: We acknowledge useful discussions with J. Allen and C. Marianetti.
This work was funded by the NNSA SSAA program through
DOE Research Grant DE-FG52-06NA26210.

\end{document}